\documentclass[onecolumn, 12pt]{IEEEtran}
\usepackage{amsmath}
\usepackage{amssymb}
\usepackage{amsfonts}
\usepackage{graphicx}
\usepackage{epsfig}
\usepackage{epstopdf}
\usepackage{subfigure}
\usepackage{psfrag}
\usepackage{cite}
\usepackage{stfloats}
\usepackage{url}
\usepackage{color}
\usepackage{bigstrut} 
\usepackage{multirow}
\usepackage{array}
\usepackage{slashbox }
\usepackage{graphicx,color,psfrag,subeqnarray}
\usepackage{latexsym}
\usepackage{authblk}
\usepackage{bm,array}
\usepackage[dvipsnames]{xcolor}
\usepackage{amsthm}
\usepackage[utf8]{inputenc}
\usepackage[english]{babel}
\renewcommand{\baselinestretch}{1.33}
\title{Real-time Shared Energy Storage  Management for Renewable Energy Integration in Smart Grid}
\author{Katayoun Rahbar, Mohammad R. Vedady Moghadam, and Sanjib Kumar Panda}
\begin{document}
\maketitle
\thispagestyle{empty}
\vspace{-4mm}
\begin{abstract}
Energy storage systems (ESSs) are essential components of the future smart grids with high penetration of renewable energy sources. 
However, deploying individual ESSs for all energy consumers, especially in large systems, may not be practically feasible mainly due to high upfront cost of purchasing many ESSs and  space limitation. 
As a result, the concept of {\it shared ESS} enabling all users charge/discharge to/from a common ESS has become appealing.  
In this paper, we study the energy management problem of a group of users with renewable energy sources and controllable (i.e., demand responsive) loads that  all share a common ESS so as to minimize their sum weighted energy cost. 
Specifically, we propose a distributed algorithm to solve the formulated problem, which iteratively derives the optimal values of charging/discharging  to/from the shared ESS, while only limited information is exchanged between users and a central controller; hence, the privacy of users is preserved. 
With the optimal charging and discharging values obtained, each user needs to independently solve a simple linear programming (LP) problem to derive the optimal energy consumption of its controllable loads over time as well as that of purchased from the grid. 
Using simulations, we show that the shared ESS can achieve lower energy cost compared to the case of distributed ESSs, where each user owns its ESS and does not share it with others. 
Next, we propose online algorithms for the real-time energy management, under non-zero prediction errors of load and renewable energy. 
The proposed algorithms differ in complexity and  the information required  to be shared between the users and central controller, where their performance is also compared via  simulations. 
\end{abstract}
\setlength{\baselineskip}{1.33\baselineskip}
\newtheorem{definition}{\underline{Definition}}[section]
\newtheorem{fact}{Fact}
\newtheorem{assumption}{Assumption}
\newtheorem{theorem}{\underline{Theorem}}[section]
\newtheorem{lemma}{\underline{Lemma}}[section]
\newtheorem{corollary}{\underline{Corollary}}[section]
\newtheorem{proposition}{\underline{Proposition}}[section]
\newtheorem{example}{\underline{Example}}[section]
\newtheorem{remark}{\underline{Remark}}[section]
\newtheorem{algorithm}{\underline{Algorithm}}[section]
\newcommand{\mv}[1]{\mbox{\boldmath{$ #1 $}}}
\begin{keywords}
	Shared energy storage system, energy management, distributed algorithm, online algorithm, renewable energy, convex optimization.
\end{keywords} 
\section{Introduction}\label{Sec:Introduction}
\PARstart{F}AST-GROWING electric energy consumption is a serious concern for existing power systems. 
According to the study reported by the US energy information administration (EIA), the worldwide energy consumption will grow by $56\%$ from 2010 to 2040 \cite{Demand}.  
An appealing sustainable solution to this concern is to widely integrate renewable energy sources into power systems and satisfy the demand of individual users locally. Besides, it helps to effectively reduce both the carbon dioxide emissions of fossil fuel based power plants and the transmission losses from power plants to the users.

The inherently stochastic nature of renewable energy generation can cause imbalanced supply and demand, which yields  fluctuations in the power system frequency and/or voltage \cite{Vedady}. 
To overcome this issue, various techniques have been proposed. 
For instance, demand response (DR) techniques adjust the power consumption of each user over time to closely match it with renewable energy generation and/or electricity prices  \cite{Vedady,DR}. 
This reduces the need of purchasing power from the grid, especially during peak-demand periods when the electricity price is high. 
However, relying solely on DR capability may not be sufficient to alleviate the stochastic renewable energy generation, since users have  must-run loads such as lighting that cannot be deferred in general. 
In this case, energy storage systems (ESSs) can be deployed  to be charged during renewable energy surplus and/or low electricity price  and be later discharged during renewable energy deficit and/or high electricity price. 

Thanks to the technology advances, integrating ESSs at the user level, e.g., residential and commercial users, is viable. 
For instance, Powerwall by Tesla \cite{Tesla} and SimpliPhi \cite{Simpliphi} have  manufactured various battery modules for residential and commercial buildings. 
However, integrating individual ESSs for all users, especially in large systems, may not be practically feasible. This is mainly due to the high upfront  cost of purchasing many ESSs (particularly in the absence of sufficient government funding) and the space limitation for installing them. 
As a result, the concept of {\it shared ESS} that enables the surplus renewable energy of some users to be charged into a shared (common) ESS and then be discharged by other users with renewable energy deficit  has become appealing in recent years \cite{Z.Wang,Paridari}. Specifically, in countries  with high population density and land scarcity  such as Singapore and Hong Kong that a large number of  users live in high-rise buildings, the concept of shared ESS is even more practically viable.    

The energy management problem for users with ESSs has been well studied in the literature. However, most of the previous works, e.g., \cite{Palomar,Katayoun_TSG_1,Katayoun_TSG_2,Zhang}, assumed that each user (or microgrid) owns an ESS that is not shared with others. The idea of sharing a common ESS among users and network operator  was introduced in \cite{Z.Wang} and interesting preliminary results were reported. 
The policy proposed in \cite{Z.Wang} for charging/discharging to/from the shared ESS and satisfying the demand responsive loads  makes decisions heuristically and only based on  the hourly electricity prices offered by the grid operator, while other practical criteria that can affect the policy of using the shared ESS are neglected. 
Beside, \cite{Paridari} solved the cost minimization problem for a system of multiple users with demand response capability and a shared ESS, but without  integration of renewable energy sources. 
The algorithm given in \cite{Paridari} aims at minimizing the total energy cost of all users, after which the total resulted benefit in cost reduction is fairly shared among all users according to their flexibility in load shifting.  
Last but not least, \cite{W.Tushar} introduced an auction based approach to capture the interaction between the shared ESS and users, where  a game theoretical technique was employed to derive the  equilibrium point of such system.

In this paper, we consider a system of multiple users each with their individually owned renewable energy generators, fixed and controllable loads, and one ESS shared among all of them. 
We assume that users exchange information with a central controller using an  existing two-way communication system, where the central controller optimizes the charging/discharging energy to/from the shared ESS by each user. 
The main contributions of this paper are  as follows:
\begin{itemize}
\item We first formulate the off-line energy management problem by assuming that the users' renewable energy generation and load are perfectly known ahead of time. We then propose an iterative based algorithm to optimally solve the off-line energy management problem in a {\it distributed} manner. The proposed algorithm requires a central controller to exchange necessary information with users so as to optimally derive  charging/discharging values to/from the shared ESS. Next, given the optimal charging and discharging values, each user independently derives the optimal energy consumption of its controllable loads and that of purchased from the grid.
\item Next, for the real-time energy management  under the practical setup of stochastic renewable energy generation/load, we deploy receding horizon control (RHC) based online algorithm \cite{MPC}, also known as model predictive control, by utilizing  the developed distributed algorithm for the off-line energy management. 
For ease of practical implementation in systems with large number of users and/or limited communication   support, we devise alternative online algorithms that are of lower complexity,  but still perform well in practice. Specifically, we propose proportional sharing (PS) and one-bit feedback (OBF) online algorithms that require the exchange of very limited amount of information at each time slot, and converge much faster than RHC. It is shown via simulations that the proposed online algorithms perform close to the optimal off-line solution, with performance gaps smaller that $7.5\%$ with about $25\%$ of renewable energy prediction errors.
\item We make comparison with benchmark case of distributed ESSs, where each user owns its relatively smaller-scale ESS, which is not shared with others. To have a fair comparison, we assume that the sum capacity of all individual ESSs in this case is equal to the capacity of the shared ESS. Our simulation results show that the shared ESS can potentially decrease the total energy cost of all users compared to the case of distributed ESSs (up to $11.5\%$ in our simulations). This is because the surplus renewable energy of one user can be utilized by others with renewable energy deficit, and also the energy curtailment is avoided more effectively due to the higher capacity of the shared ESS. We further highlight the impact of renewable energy sources diversity on the effectiveness of shared ESS in energy cost reduction.
\end{itemize}

In contrast to the prior works \cite{Zhang,Palomar,Katayoun_TSG_1,Katayoun_TSG_2,Z.Wang,Paridari,W.Tushar}, we propose a distributed algorithm for the  energy management of users with a shared ESS.  Our recent works on shared ESS management in \cite{Katayoun_Shared_1,Katayoun_Shared_2} provide preliminary results on the effectiveness of shared ESS in energy cost reduction \cite{Katayoun_Shared_1} and how self-interested users can trade energy with the shared ESS to simultaneously achieve lower energy costs while preserving their privacy \cite{Katayoun_Shared_2}. In contrast to \cite{Katayoun_Shared_1,Katayoun_Shared_2}, in this paper, we investigate online algorithms for the real-time energy management of a system constituting of a shared ESS, which to the best of our knowledge has not been rigorously studied yet. 
Note that the real-time energy management for users with a shared ESS highly differs from that of distributed ESSs and/or no ESS integration that has been comprehensively studied in the  literature, e.g., see  \cite{Katayoun_TSG_2,Online_1,Online_2}, since users are all coupled together through the shared ESS.

The rest of this paper is organized as follows. Section \ref{Sec:SysModel} presents the system model and formulates the optimization problem. Section \ref{Sec:Offline} presents a distributed algorithm to   optimally solve the off-line energy management problem and also provides a benchmark setup of users with distributed ESSs. Section \ref{Sec:Online} presents three online
algorithms for the real-time energy management. Section \ref{Sec:Simul} presents simulation results. Last, we conclude the paper in Section \ref{Sec_Concl}.  
\section{System Model and Problem Formulation}\label{Sec:SysModel}
\begin{figure}[t!]
	\centering
	\includegraphics[width=8cm]{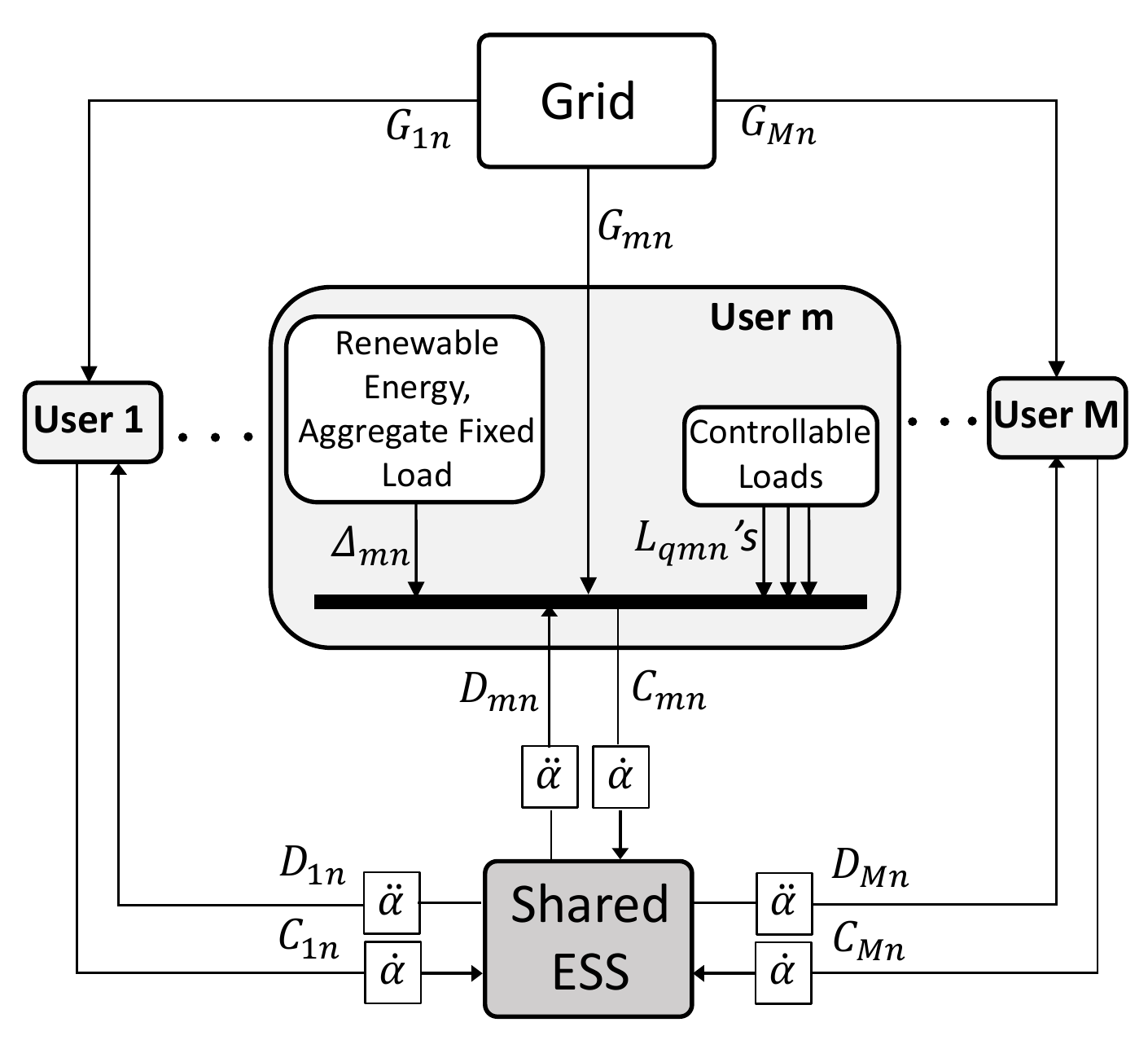}  
	\caption{System model: Users with a shared ESS.}\label{fig:System_Model} 
\end{figure}

As shown in Fig. \ref{fig:System_Model}, we consider a system of $M > 1$ users, indexed by $m$, $m \in {\cal M} = \{1,\ldots,M\}$, each of which can be a single residential, commercial, or industrial  energy consumer  or even  a group of consumers managed by an aggregator. 
We consider that each user has its own renewable energy generator  that supplies a part or all of its demand over time, but it is still grid-connected and can draw energy from the grid whenever necessary.  
We also consider that each user has two different types of loads, namely {\it fixed} and {\it controllable}, where each fixed load (e.g., lighting) should be instantly satisfied upon the request of the user, while each controllable load (e.g., smart electric water heater) can be satisfied  within a desired time period specified by the user a priori, subject to certain  practical considerations. 
Furthermore, we consider that there is an energy storage system (ESS) shared among all users, where the surplus energy of some users can be charged into it and be then discharged  by others with renewable energy deficit at the same time and/or later.  
A central controller is then assumed for coordinating the use of the shared ESS by exchanging the required information with  users  through an existing bidirectional communication system. 
Specifically, the central control unit is responsible for {\it joint} optimization of energy charged/discharged to/from the shared ESS by all users to minimize their sum weighted energy cost. Last, for convenience, we adopt a time-slotted system with slot index $n$, $n \in {\cal N}=\{1,\ldots,N\}$, with $N \ge 1$ denoting the total number of scheduling times slots, where the duration of each slot is normalized to a unit time, hence power and energy are used interchangeably in this paper.
In the following, we define the system model in  detail.
\subsubsection{Grid Energy Cost} Let $G_{mn} \ge 0$ denote the energy drawn from the distribution company/grid by user $m$ at time slot $n$, where the corresponding cost for the user is modeled by $f_{mn}(G_{mn}) \ge 0$. We assume that $f_{mn}(\cdot)$, which is time-varying in general, is a convex and monotonically increasing function over $G_{mn} \ge 0$. 

\subsubsection{Shared ESS} Let $C_{mn}$ and $D_{mn}$, with $0 \le C_{mn} \le \overline{C}$ and $0 \le D_{mn} \le \overline{D}$, denote the energy charged/discharged to/from the shared ESS by  user $m$ at time slot $n$, respectively, where $\overline{C}>0$ and $\overline{D}>0$ are the given maximum charging and discharging rates of the shared ESS per user, respectively.  
The energy losses during  charging and discharging processes are specified by charging and discharging efficiency parameters, denoted by $0< \dot{\alpha} < 1$ and $0 <\ddot{\alpha} < 1$, respectively. Denote $S_n\ge 0$ as the available energy in the shared ESS at the beginning of time slot $n$,  which can be derived recursively from the state,  charging, and discharging values of all users in the previous time slot as follows: 
\begin{align} \label{eq:storage-State}
S_{n+1}=S_n+\dot{\alpha} \sum_{m=1}^{M}{C_{mn}} - \frac{1}{\ddot{\alpha}} \sum_{m=1}^{M}{D_{mn}}.
\end{align}
A practical ESS  has a finite capacity  and cannot be  fully discharged to avoid deep discharging. 
Hence, we consider the following constraints for the states of the shared ESS: 
\begin{align}
~&S_1+\sum_{m=1}^{M}\sum_{i=1}^{n}(\dot{\alpha}C_{mi}-1/\ddot{\alpha}{D}_{mi}) \ge \underline{S}, \forall n\in \cal N \label{P1:Const1}\\
~&S_1+\sum_{m=1}^{M}\sum_{i=1}^{n}(\dot{\alpha}C_{mi}-1/\ddot{\alpha}{D}_{mi}) \le \overline{S}, \forall n\in \cal N \label{P1:Const2}
\end{align}
where $\underline{S} > 0$ and $\overline{S} > \underline{S}$ are the minimum and maximum allowable states of the shared ESS, respectively. We set $\underline{S} \le S_{1} \le \overline{S}$ by default.

\subsubsection{Controllable and Fixed Loads} Denote $\hat{L}_{mn} \ge 0$ as the aggregate fixed loads of user $m$ at time slot $n$.  
We assume that each user  has multiple controllable loads whose energy consumption can be scheduled flexibly, discussed as follows. 
Let $Q_m\ge 1$ denote the number of controllable loads of user $m$, indexed by $q$, $q \in {\cal Q}_m=\{1,\ldots,Q_m\}$. 
Specifically, we consider that controllable load $q$ of user $m$ requires $E_{qm}>0$ amount of energy to complete its task  over  time slots $\underline{n}_{qm}\le n \le \overline{n}_{qm}$, where $1\le \underline{n}_{qm}<N$  and  $\underline{n}_{qm}< \overline{n}_{qm}\le N$ are the start and termination time slots, respectively, which are specified by user $m$.  
Accordingly, we define ${\cal N}_ {qm}=\{\underline{n}_{qm},\ldots,\overline{n}_{qm} \}$. 
Let $L_{qmn} \ge 0$ denote the energy allocated to controllable load $q$ of user $m$ at time slot $n$. Due to practical considerations, $L_{qmn}$ should be higher (lower) than a given minimum (maximum) threshold $\underline{L}_{qm} \ge 0$ ($\overline{L}_{qm}> \underline{L}_{qm}$)  over time slots $n \in {\cal N}_{qm}$. 
Without loss of generality, we set  $\underline{L}_{qm}=\overline{L}_{qm}=0$, $\forall n \not \in {\cal N}_{qm}$.  
By default, we set $(\overline{n}_{qm}-\underline{n}_{qm})\underline{L}_{qm} <E_{qm} < (\overline{n}_{qm}-\underline{n}_{qm})\overline{L}_{qm}$, to ensure that controllable load $q$ of user $m$ is  practically schedulable.\footnote{We assume that the energy consumption of demand responsive loads can be {\it continuously} changed over time which accurately models  electric water heaters, heating, and cooling systems, that account for the largest energy consumption of residential consumers.}   
We thus have the following constraints for controllable loads:  \vspace{-2mm}
\begin{align}
&\sum_{n=\underline{n}_{qm}}^{\overline{n}_{qm}}{{L}_{qmn}}=E_{qm},~\forall q \in {\cal Q}_m,~\forall m \in \cal M,\label{P1:Const3}\\
&\underline{L}_{qm} \le {L}_{qmn} \le \overline{L}_{qm},~\forall q \in {\cal Q}_m,~{\forall m} \in {\cal M},~\forall n \in \cal N. \label{P1:Const4}
\end{align}
 
\subsubsection{Energy Neutralization} By denoting $R_{mn} \ge 0$ as the renewable energy generation of user $m$ at time slot $n$, we define ${\Delta}_{mn}=R_{mn}-\hat{L}_{mn}$ as renewable energy generation offset by the aggregate fixed load, where ${\Delta}_{mn} \ge 0$ indicates that renewable energy generation can fulfill the aggregate fixed load of user $m$ at time slot $n$ and may also meet all or part of its controllable load demand and/or charge the shared ESS.  
On the other hand, $\Delta_{mn}<0$ indicates that the generated renewable energy of user $m$ at time slot $n$ cannot even meet its aggregate fixed load. 
We assume that renewable energy generation and fixed load  of each user are  practically predictable but with finite prediction errors.\footnote{We assume that there is no randomness in starting/termination times of the controllable loads and their total energy consumption throughout their scheduling period.} 
By denoting $\overline{\Delta}_{mn}$ as the predictable net energy profile, we have 
${\Delta}_{mn}= \overline{\Delta}_{mn}+\delta_{mn}$, where ${\delta}_{mn}$ denotes  prediction errors.
Finally, we assume that each user $m$ needs to satisfy its fixed and controllable loads at each time slot $n$, using its renewable energy generation, shared ESS, and/or purchasing from the grid. We then have the following energy neutralization constraints for all users:\vspace{-.2cm}
\begin{align}
G_{mn}\hspace{-.7mm}-\hspace{-.7mm}C_{mn}\hspace{-.7mm}+\hspace{-.7mm}D_{mn}\hspace{-.7mm}+\hspace{-.7mm}{\Delta}_{mn}\hspace{-.7mm}-\hspace{-.7mm}\sum_{q=1}^{Q_m}\hspace{-.7mm}{L_{qmn}} \hspace{-.3mm}\ge 0, {\forall m} \in {\cal M}, \forall n \in \cal N. \label{P1:Const6}
\end{align}

With the system model defined  above, we now proceed to minimize the sum weighted energy cost of all users, subject to practical constraints of their loads as well as the shared ESS. 
We thus formulate the following optimization problem.\vspace{-1mm}
\begin{align}
\mathrm{(P1)}:  \mathop{\mathtt{min}}_{\{{\cal  X}_m\}_{\forall m \in {\cal M}}}~& \sum_{m=1}^{M} \sum_{n=1}^{N}\beta_m f_{mn}({G}_{mn}) \nonumber\\ 
\mathtt{s.t.} 
~&(\ref{P1:Const1})-(\ref{P1:Const6}),\nonumber
\end{align}
where ${\cal X}_m \triangleq \{0 \le C_{mn} \le \overline{C},0 \le D_{mn} \le \overline{D},G_{mn} \ge 0,L_{qmn},\forall q \in {\cal Q}_m, \forall n\in \cal N\}$ denotes the set of all decision variables for user $m$. Moreover, $\beta_m$'s, with $0 \le \beta_m \le 1$, $\forall m \in \cal M$ and $\sum_{m=1}^{M}\beta_m=1$, are the given cost weight coefficients for different users.\footnote{In practice, there are different approaches to set the cost weight coefficients  $\beta_m$'s. For instance, consider the scenario that users invest to purchase a bulk ESS. In this case, $\beta_{mn}$'s can be set such that users benefit from the shared ESS according to their initial investment.}\vspace{-.8mm}
\section{Shared ESS Management: Off-line Optimization}\label{Sec:Offline}
In this section, we first propose an algorithm to optimally solve (P1) in a distributed manner, by  assuming that $\Delta_{mn}$'s are perfectly known to each user without any prediction errors. To do so, we use the so-called  duality principle. 
We also formulate the benchmark case of distributed ESSs, where users have their individually owned ESS that is not shared with others.\vspace{-.8mm}
\subsection{Distributed Algorithm for (P1)}
For convenience, we introduce vector presentation for decision variables in (P1) as well as some system parameters as   $\mv{g}_{m}=[G_{m1},\ldots,G_{mN}]^T$, $\mv{c}_{m}=[C_{m1},\ldots,C_{mN}]^T$, $\mv{d}_{m}=[D_{m1},\ldots,D_{mN}]^T$,  $\mv{l}_{qm}=[L_{qm1},\ldots,L_{qmN}]^T$,  $\hat{\mv{l}}_{m}=[\hat{L}_{m1},\ldots,\hat{L}_{mN}]^T$,  $\mv{r}_{m}=[R_{m1},\ldots,R_{mN}]^T$, and 
 $\mv{\Delta}_{m}=\mv{r}_{m}-\hat{\mv{l}}_{m}=[\Delta_{m1},\ldots,\Delta_{mN}]^T$. 
Moreover, let $\mv{y}_m=[Y_{m1},\cdots,Y_{mN}]^T$, with $Y_{mn} \ge 0$, $\forall n \in \cal N$, denote the Lagrange variables corresponding to energy neutralization constraints in (\ref{P1:Const6}). The Lagrangian of (P1) is thus expressed as\vspace{-1mm}
\begin{align}
&\mathcal{L}(\{{\cal X}_m\}_{\forall m \in \cal M},\{\mv{y}_m\}_{\forall m \in \cal M})=\sum_{m=1}^{M}\sum_{n=1}^{N}{\beta_m f_{mn}(G_{mn})} -\sum_{m=1}^{M}\mv{y}_m^T(\mv{g}_{m}-\mv{c}_{m}+\mv{d}_{m}+\mv{\Delta}_{m}-\sum_{q=1}^{Q_m}{\mv{l}_{qm}}).
\end{align}
The dual function of $\mathcal{L}(\cdot)$ is given by \vspace{-1mm}
\begin{align}\label{Dual}
g(\{\mv{y}_m\}_{\forall m \in \cal M})=  \mathop{\mathtt{min}}_{\{{\cal  X}_m\}_{\forall m \in \cal M}}&~ \mathcal{L}(\{{\cal X}_m\}_{\forall m \in \cal M},\{\mv{y}_m\}_{\forall m \in \cal M})\nonumber\\ 
 \mathtt{s.t.}&~(\ref{P1:Const1})-(\ref{P1:Const4}).
\end{align}
Hence, the dual problem of (P1) is derived as \vspace{-1mm}
\begin{align}
\mathrm{(D1)}: \mathop{\mathtt{max}}_{\{\mv{y}_m \ge 0\}_{\forall m\in \cal M}}~ g(\{\mv{y}_m\}_{\forall m\in \cal M}).
\end{align}
Since (P1) is a convex optimization problem and satisfies the Slater's condition \cite{Boyd}, strong duality holds between (P1) and (D1). 
Therefore, (P1) can be solved by investigating the optimal solution to its dual problem (D1) equivalently. 
To solve (D1), we use the \textit{subgradient} method \cite{Subgradient}, which can be implemented via an iterative algorithm as follows.

Let  $\{{\cal  X}_m^{(k)}\}_{\forall m \in \cal M}$, with ${\cal  X}_m^{(k)}=\{\mv{c}_m^{(k)}, \mv{d}_m^{(k)},\mv{g}_m^{(k)},\mv{l}_{qm}^{(k)},\forall q \in {\cal Q}_m\}$, and $\{\mv{y}_m^{(k)}\}_{\forall m \in \cal M}$ denote the values of primal and dual (Lagrange) variables of (P1), respectively, at each iteration $k$, $k=1,2,\cdots$.  
At iteration $k$,  (\ref{Dual}) is firstly solved with fixed   $\{\mv{y}_m^{(k-1)}\}_{\forall m \in \cal M}$, where 
$\{\mv{y}_m^{(0)} \ge 0\}_{\forall m \in \cal M}$ denotes an initial point that is  randomly generated.  
Hence, (\ref{Dual}) can be decoupled into the following sub-problems:  
\begin{align} 
\mathop{\mathtt{min}}_{\mv{g}_m \ge 0, \{\mv{l}_{qm}\}_{\forall q \in {\cal Q}_m}} &\sum_{n=1}^{N}{\hspace{-0.5mm}\beta_m f_{mn}(G_{mn})}\hspace{-0.5mm}-\hspace{-0.5mm}\mv{y}_m^{(k-1)\hspace{1mm}T}\hspace{-0.5mm}(\mv{g}_m\hspace{-0.8mm}-\hspace{-0.8mm}\sum_{q=1}^{Q_m}{\mv{l}_{qm}})\nonumber\\
\mathtt{s.t.}&~(\ref{P1:Const3})\text{ and }(\ref{P1:Const4}),\label{Sub_1} 
\end{align}
for $m=1,\ldots,M$, and\vspace{-1mm}
\begin{align}
\mathop{\mathtt{min}}_{\{0 \le \mv{c}_{m} \le \overline{C}\}_{\forall m \in \cal M}, \{0 \le \mv{d}_{m} \le \overline{D}\}_{\forall m \in \cal M}} &\sum_{m=1}^{M}{\mv{y}_m^{(k-1)\hspace{1mm}T}(\mv{c}_m-\mv{d}_m)}\nonumber \\
\mathtt{s.t.}&~(\ref{P1:Const1})-(\ref{P1:Const2}).\label{Sub_2}
\end{align}
For each $m$, $\mv{g}_m^{(k)}$ and $\{\mv{l}_{qm}^{(k)}\}_{q\in {\cal Q}_m}$ are set as the optimal solution to (\ref{Sub_1}), which is unique due to the fact that every cost function $f_{mn}(G_{mn})$ is assumed to be  convex and monotonically increasing  over $G_{mn}\ge 0$. However, the optimal solution to the linear programming (LP) in (\ref{Sub_2}) is generally not unique. In this case, to ensure the smooth convergence of algorithm, the so-called running average technique \cite{Averaging_1,Averaging_2} is used, under which $
\mv{c}_m^{(k)} = 1/k\sum_{i=1}^{k}\hat{\mv{c}}_m^{(i)}$ and $\mv{d}_m^{(k)} = 1/k\sum_{i=1}^{k}\hat{\mv{d}}_m^{(i)}$, $\forall m \in \cal M$, with $\{\hat{\mv{c}}_m^{(k)}\}_{\forall m \in \cal M}$ and $\{\hat{\mv{d}}_m^{(k)}\}_{\forall m \in \cal M}$ denoting any  optimal solution  to (\ref{Sub_2}).  
With $\{{\cal  X}_m^{(k)}\}_{\forall m \in \cal M}$ obtained as above, the dual function  $g(\{\mv{y}_m\}_{\forall m \in \cal M})$ in (\ref{Dual}) is  formed, where it can be shown that the resulted  dual function is concave  but not necessarily differentiable \cite{Boyd}.  
Nevertheless, it can be verified that the subgradient of  $g(\{\mv{y}_m\}_{\forall m \in \cal M})$ always exists \cite{Subgradient}, where the subgradient corresponding to  user $m$ at iteration $k$ is given by 
\begin{align} \label{vm}
\mv{v}_m^{(k)}=\mv{z}_{m}^{(k)}-\mv{c}_{m}^{(k)}+\mv{d}_{m}^{(k)},
\end{align} 
with  
\begin{align} \label{zm}
\mv{z}_m^{(k)}=\mv{g}_m^{(k)}-\sum_{q=1}^{Q_m}{\mv{l}_{qm}^{(k)}}+\mv{\Delta}_m.
\end{align}
Accordingly,  the dual variables can be updated  by using  subgradient based methods such as the ellipsoid method \cite{Subgradient}, where the updated dual variables are denoted by   $\{\mv{y}_m^{(k)}\}_{\forall m \in \cal M}$. 
The algorithm terminates when the dual variables all converge within a prescribed accuracy. 
Let $\{{\cal  X}_m^{\star}\}_{\forall m}$, with ${\cal  X}_m^{\star}=\{\mv{c}_m^{\star},\mv{d}_m^{\star},\mv{g}_m^{\star},\mv{l}_{qm}^{\star},\forall q \in {\cal Q}_m\}$, and $\{\mv{y}_m^{\star}\}_{\forall m}$  denote the optimal primal and dual variables of (P1), respectively. 
\begin{table}[t!]
	\begin{center}
		\caption{Algorithm for the optimal off-line energy management}
		\vspace{0.01cm}
		\hrule
		\vspace{0.05cm} \textbf{Algorithm 1}  \vspace{0.03cm}
		\hrule \vspace{0.01cm}
		\begin{itemize}
			\item[1)] Initialize $\{\mv{y}_m^{(0)} \ge {0}\}_{\forall m \in \cal M}$. 
			\item[2)] {\bf Repeat:}
			\begin{itemize}
				\item[a)]  Given   $\mv{y}_m^{(k-1)}$ received from the central controller, user $m$ solves  (\ref{Sub_1}) and saves the obtained solution as  $\mv{g}_m^{(k)}$ and $\{\mv{l}_{qm}^{(k)}\}_{\forall q \in {\cal Q}_m}$. Accordingly, user $m$ evaluates  $\mv{z}_m^{(k)}$ in (\ref{zm}), and sends it to the central controller via the existing communication system;
				\item[b)]  Given $\{\mv{y}_m^{(k-1)}\}_{\forall m \in \cal M}$, the central controller solves (\ref{Sub_2}), and then by employing the averaging technique \cite{Averaging_1,Averaging_2} derives  $\{\mv{c}_m^{(k)}\}_{\forall m}$ and $\{\mv{d}_m^{(k)}\}_{\forall m}$. Furthermore, the central controller evaluates the subgradients $\mv{v}_m^{(k)}$, $\forall m \in \cal M$, in (\ref{vm}), and accordingly updates the dual variable via e.g.  the ellipsoid method \cite{Subgradient}, where the updated dual variables are saved as $\{\mv{y}_m^{(k)}\}_{\forall m \in \cal M}$. 
			\end{itemize}
			\item[3)]  {\bf Until}  the dual variables all convergence within a prescribed accuracy. 
			\item[4)] The central controller sets  $\mv{y}_m^{\star}\leftarrow \mv{y}_m^{(k)} $, $\forall m \in \cal M$, and broadcasts $\mv{c}_m^{(k)}$ and $\mv{d}_m^{(k)}$ to each user $m$.
			\item[5)] Each user $m$ sets $\mv{c}_m^{\star}\gets \mv{c}_m^{(k)}$ and $\mv{d}_m^{\star}\gets \mv{d}_m^{(k)}$,  $\mv{g}_{m}^{\star}\leftarrow\mv{g}_{m}^{(k)}$, and $\{\mv{l}_{qm}^{\star}\}_{\forall q \in {\cal Q}_m} \leftarrow \{\mv{l}_{qm}^{(k)}\}_{\forall q \in {\cal Q}_m}$. 
		\end{itemize}
		\vspace{0cm} \hrule \vspace{0.2cm}\label{Algorithm:1}
	\end{center} 	
\end{table}
The aforementioned  algorithm to optimally solve (P1) is thus summarized in Table \ref{Algorithm:1}, as Algorithm $1$. 

It is worth noting that at each iteration of Algorithm $1$, the updates in (\ref{Sub_1})--(\ref{zm}) can be implemented in a fully distributed manner, discussed as follows. 
First, given $\mv{y}_m^{(k-1)}$ announced by the central controller to each user $m$, the user derives  $\mv{g}_m^{(k)}$ and $\{\mv{l}_{qm}^{(k)}\}_{\forall q \in {\cal Q}_m}$ by solving (\ref{Sub_1}) independently, after which it computes $\mv{z}_m^{(k)}$ in (\ref{zm}) and sends the obtained value to the central controller via the existing communication system.
On the other hand, the central controller  obtains $\{\mv{c}_m^{(k)}\}_{\forall m \in \cal M}$ and $\{\mv{d}_m^{(k)}\}_{\forall m \in \cal M}$ by  solving (\ref{Sub_2}) and also applying  the averaging technique \cite{Averaging_1,Averaging_2}.   
Moreover,  the central controller computes the subgradients $\mv{v}_m^{(k)}$, $m=1,\ldots,M$, in (\ref{vm}) and updates the dual variables via e.g. the ellipsoid method \cite{Subgradient}, where the updated dual variables are saved as $\{\mv{y}_m^{(k)}\}_{\forall m \in \cal M}$.  
\begin{remark}
Algorithm $1$ preserves the  privacy of users, since at each iteration,  user $m$ only needs to share  $\mv{z}_m^{(k)}$ in (\ref{zm}) with the central controller. Hence, the user do not need to share the detailed information of its load characteristics, renewable energy profile, etc. with the central controller and/or other users.  
\end{remark} 
\subsection{Benchmark: Distributed ESSs}\label{Sec:SharedvsDist}
In this subsection, we consider a system of distributed ESSs, where users own their individual small-scale ESSs that are not shared with any other user. 
\begin{figure}[t!]
	\centering
	\includegraphics[width=8.25cm]{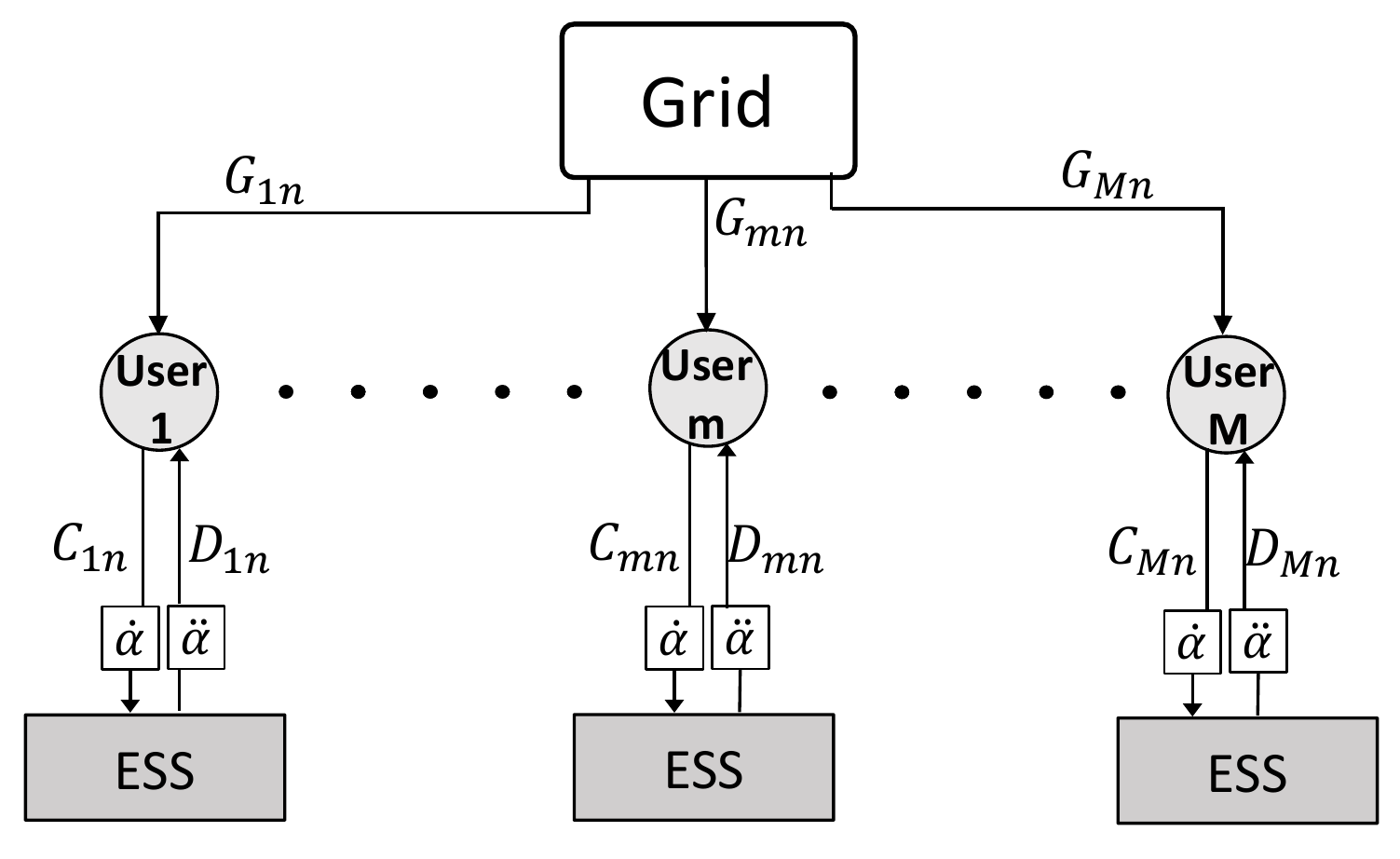} 
	\caption{System model: Users with distributed ESSs.}\label{fig:Dist_ESS} 
\end{figure}
In this case, the state of
the ESS for each user $m$, denoted by $S_{mn}$ at time slot $n$, is given by 
\begin{align}
S_{mn+1}=S_{mn}+\dot{\alpha}_m{C_{mn}} - \frac{1}{\ddot{\alpha}d_m} {D_{mn}}, \label{Storage_Split}
\end{align}
where $ 0<\dot{\alpha}_m<1$ and $0<\ddot{\alpha}_m<1$ are charging and discharging efficiency parameters, respectively. Similar to the shared ESS, we have the following constraints: 
\begin{align}\label{eq:storage_distributed}
\underline{S}_m \le S_{mn} \le \overline{S}_m,~\forall n \in {\cal N}, 
\end{align}
where $\underline{S}_m\ge 0$ and $\overline{S}_m\ge 0$ are the minimum and maximum allowed states of the ESS owned by user $m$. 
The charging and discharging values should satisfy $0 \le C_{mn} \le \overline{C}_m$ and $0 \le D_{mn} \le \overline{D}_m$, where  $\overline{C}_m>0$ and $\overline{D}_m>0$ are maximum charging and discharging rates, respectively.  
To have a fair comparison with the case of shared ESS,  we set $\underline{S}=\sum_{m=1}^{M}\underline{S}_m$, $\overline{S}=\sum_{m=1}^{M}\overline{S}_m$, $\overline{C}=\sum_{m=1}^{M}\overline{C}_m$, and $\overline{D}=\sum_{m=1}^{M}\overline{D}_m$. We also set $\dot{\alpha}_m=\dot{\alpha}$ and $\ddot{\alpha}_m=\ddot{\alpha}$, $\forall m \in \cal M$.

We now formulate the optimization problem as follows. 
\begin{align*}
\mathrm{(P2)}:  \mathop{\mathtt{min}}_{{\{{\cal  X}_m\}}_{\forall m}}~& \sum_{m=1}^{M}\sum_{n=1}^{N}\beta_m f_{mn}(G_{mn}) \nonumber\\ 
\mathtt{s.t.} 
~&\underline{S}_m \le S_{mn} \le \overline{S}_m,~\forall m \in {\cal M},~\forall n \in {\cal N}\\
~&(\ref{P1:Const3})-(\ref{P1:Const6}).
\end{align*}
It can be readily verified that (P2) is convex \cite{Boyd} and  separable over all users, since ESS constraints are not coupled over users.  In this case, (P2) can be decomposed into $M$ subproblems, one for each user, and be solved without the need of information exchange between users and the central controller. 
\section{Online Energy Management}\label{Sec:Online}
In Section \ref{Sec:Offline}, we have discussed an off-line scenario where the  renewable energy generation and energy consumption of (fixed) loads of the users are all known {\it a priori}, i.e., predicted without error.  However, this assumption does not hold in practice, even by using the most advanced forecasting techniques. As a result, in the following, we consider three online algorithms for the real-time energy management of the system with non-zero prediction errors, i.e., $\delta_{mn}\neq 0$, $\forall m\in \cal M$, $\forall n \in \cal N$. 
\subsection{Receding Horizon Control (RHC) Based Online Algorithm} 
In this algorithm, starting from time slot $n=1$ to $n=N$,  the energy management problem is solved via Algorithm $1$ over windows with receding sizes of $N-n-1$. 
Specifically, at each time slot $n$,  actual values of renewable energy generation/load of past and current time slots, i.e., $1,\ldots,n$, and the predictable values of the future i.e., $n+1,\ldots,N$, are used  in Algorithm $1$ to derive the decision variables of the current time slot $n$, i.e.,  $\{C_{mn},D_{mn},G_{mn},L_{qmn},\forall q \in {\cal Q}_m, \forall m\in \cal M\}$.  
Please refer to \cite{Katayoun_TSG_1,MPC} for more detail of RHC algorithm. 

Although RHC is a conventional technique for real-time energy management and has a close-to-optimal performance in general, its implementation for systems with large number of users ($N\gg1$) and/or limited communication support is challenging due to the high computational complexity and large amount of information exchange between users and the central controller.
For ease of practical implementation, in the following, we devise two alternative online algorithms of low complexity that require limited information sharing between users and the central controller, have relatively faster convergence rates, and also perform close to the optimal off-line solution derived by assuming zero  prediction errors.  
\subsection{Proportional Sharing (PS)  Online Algorithm}
At each time slot $n$,  starting from user $1$ to $M$, if user $m$ has energy deficit, it  announces its \textit{modified} net energy profile to the central controller, where the  modified net energy profile will be discussed later in the following.  
Accordingly, the central controller  forms a set, denoted by $\cal M^D$, whose elements are users with energy deficit. 
On the other hand, users with energy surplus firstly satisfy their fixed and controllable loads as much as possible, but subject to their constraints. 
Then, they send their surplus renewable energy (if any) to the shared ESS (either to be stored in the shared ESS and/or curtailed when the ESS is full). 
Next, the central controller {\it proportionally} divides the available energy in the shared ESS among users $m\in {\cal M^D}$ based on their energy deficit feedbacked. 
The remaining energy deficit of users (if any) is finally  satisfied from the grid. 

\begin{figure}[t!]
	\centering
	\includegraphics[width=10.5cm]{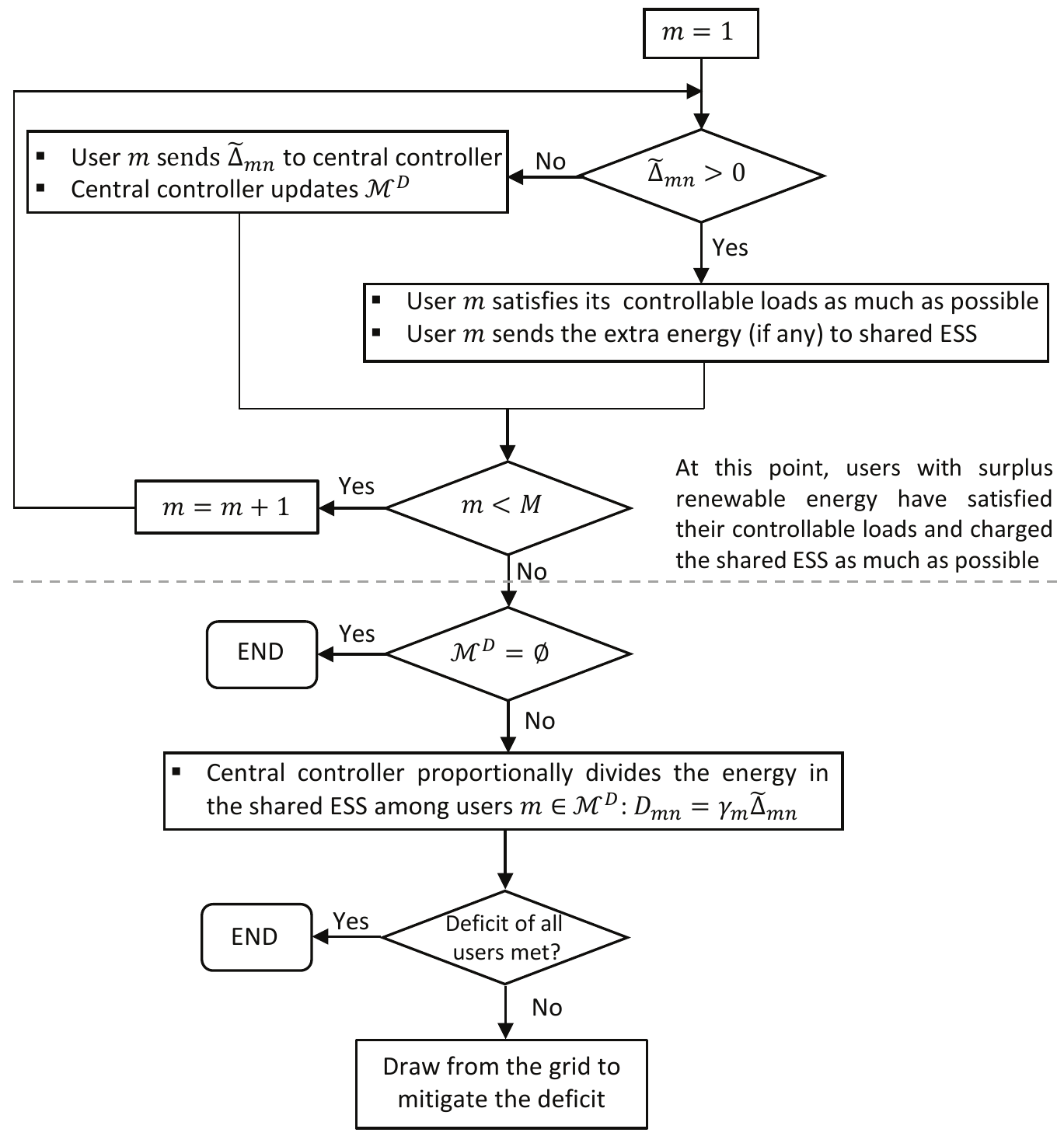} 
	\caption{PS online algorithm at each time slot $n$.}\label{fig:One-bit} 
\end{figure}

The PS based online algorithm makes decisions based on the instantaneous energy surplus/deficit of users and the available energy in the shared ESS, hence it needs to ensure that the total energy requirements of controllable loads of all users are met by their given  termination time slots.
To do so, at time slot $n$, each user $m$ firstly needs to evaluate   ${\tilde{L}_{qmn}}=[E_{qm}-\sum_{i=1}^{n-1}{L_{qmn}}]^+/(\overline{n}_{qm}-n+1)$, with  $\underline{L}_{qm}\le \tilde{L}_{qmn} \le \overline{L}_{qm}$, which  shows the unsatisfied energy  of its controllable load $q$  normalized over the number of time slots left to reach its termination time slot $\overline{n}_{qm}$.  
Hence, if ${\tilde{L}_{qmn}}$ amount of energy  is assigned to this controllable load over time slots $n,\ldots,\overline{n}_{qm}$, its energy requirement will be met surely.
Accordingly, user $m$ can set its {\it modified} fixed load as  $\tilde{L}_{mn}=\hat{L}_{mn}+\sum_{q=1}^{Q_m} \tilde{L}_{qmn}$. 
The {\it modified} net energy profile for user $m$ at time slot $n$ is thus obtained as  $\tilde{\Delta}_{mn}=R_{mn}-\tilde{L}_{mn}$. 
With such modified net energy profiles, the central controller proportionaly divides the energy in the shared ESS among users $m \in {\cal M}^D$ as $D_{mn}= \gamma_m \tilde{\Delta}_{mn} $, where $ \gamma_m=\min\{\ddot{\alpha}S_n,\overline{D}\}/\sum_{m\in {\cal M}^D}{\tilde{\Delta}_{mn}}$. 
To summarize, the aforementioned PS  online algorithm is presented in Fig. \ref{fig:One-bit}. 
\subsection{One-Bit Feedback (OBF) Online Algorithm}
This algorithm performs the same as that in PS online algorithm except that at each time slot $n$, each user $m$ after calculating its modified net energy profile $\tilde{\Delta}_{mn}$, sends only one bit feedback to the central controller to indicate its energy surplus/deficit (no need to send the exact value of surplus/deficit). For instance, by sending $1$ when $\tilde{\Delta}_{mn}<0$, the central controller is notified of energy deficit, while $0$ is sent when $\tilde{\Delta}_{mn} \ge 0$, indicating energy surplus or zero net energy profile. 
Due to receiving only one bit of information, the central controller {\it evenly} divides the available energy in the shared ESS among all users with energy deficit, i.e., $m\in {\cal M}^D$, regardless of the amount of their deficit.

Last, note that RHC based online algorithm uses an iterative algorithm given in Table \ref{Algorithm:1}  at each time slot $n$, which requires the large amount of information exchange between the central controller and different users. However,  PS and OBF  online algorithms require the exchange of very limited amount of information at each time slot and converge  faster; hence, can be implemented in systems with large number of users and/or limited communication support. However, the performance of PS and OBF online algorithms is expected to degrade compared to RHC based online algorithm, as will be shown in Section \ref{SubSec:Online}. 
\section{Simulation Results}\label{Sec:Simul}
\begin{figure}[t!]
	\centering
	\includegraphics[width=12cm]{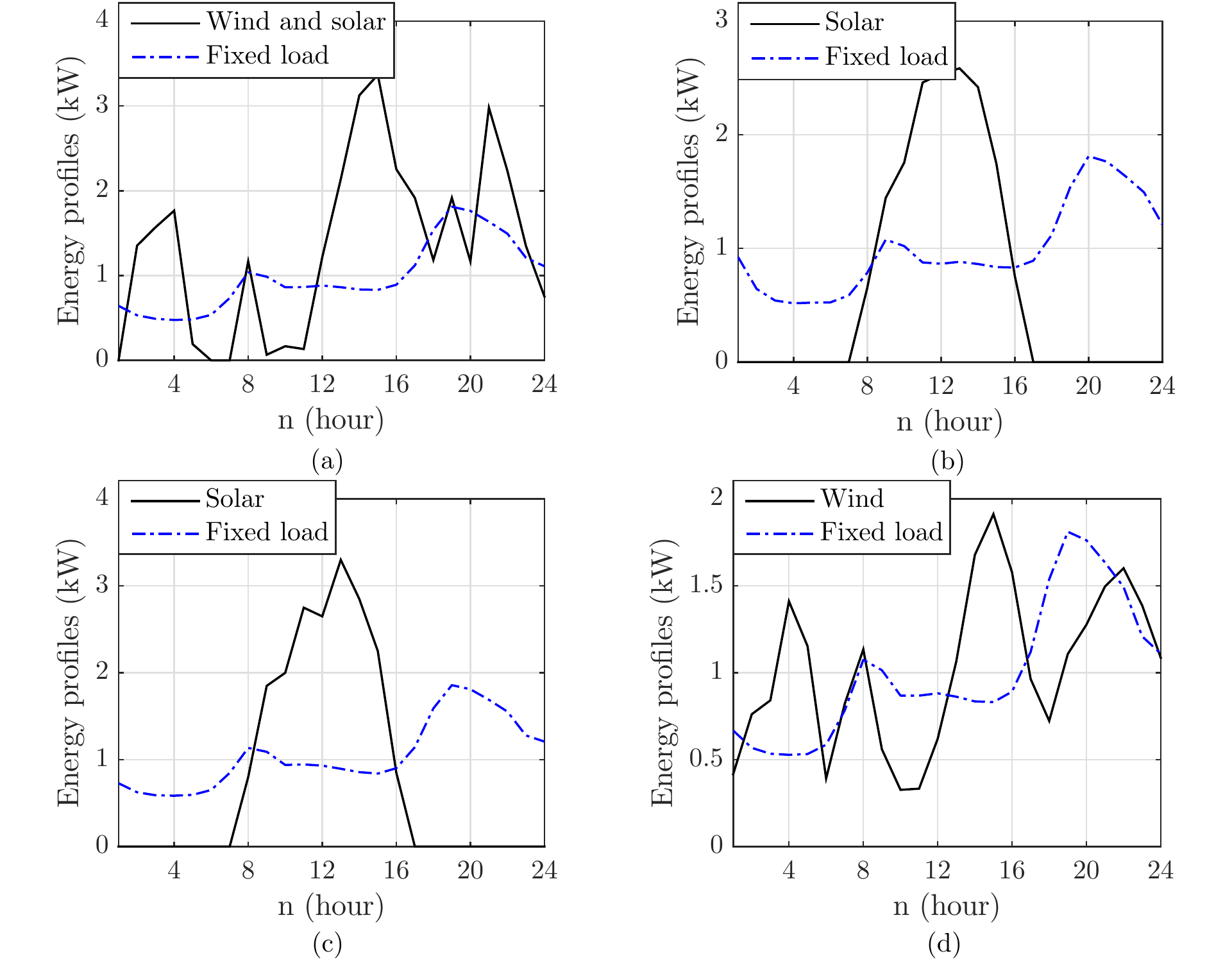} 
	\caption{Energy profiles of renewable energy generation and fixed loads of: (a) user 1, (b) user 2, (c) user 3, and (d) user 4.}\label{fig:Profiles}  
\end{figure}

We consider a system of four residential users $M=4$, each integrating  its renewable energy generators, solar and/or wind, over one day $N=24$. We consider that $n=1$ indicates time 00:00 AM, $n=2$ time 01:00 AM, and finally $n=24$ time 11:00 PM.  
Energy profiles of renewable energy generation and users' fixed loads are shown in Fig. \ref{fig:Profiles} \cite{NREL,NREL_Solar,Load_Profile}.
\begin{table}[t!]
	\centering
	\caption{Controllable loads parameters} 
	\label{Table:Flexible_Load}
	\begin{tabular}{|c|c|c|c|c|c|}
		\hline
		\multirow{2}{*}{\begin{tabular}[c]{@{}c@{}}Controllable \\ Loads\end{tabular}}       & \multirow{2}{*}{$\underline{n}_{qm}$\hspace{-.5mm}} & \multirow{2}{*}{$\overline{n}_{qm}$\hspace{-.5mm}} & \multirow{2}{*}{$\underline{L}_{qm}$\hspace{-.6mm} (kW)\hspace{-1mm}} & \multirow{2}{*}{$\overline{L}_{qm}$\hspace{-.6mm} (kW)\hspace{-1mm}} & \multirow{2}{*}{$E_{qm}$\hspace{-1mm} (kWh)\hspace{-1mm}} \\
		&                                       &                                      &                                            &                                           &                                 \\ \hline
		\multirow{2}{*}{\begin{tabular}[c]{@{}c@{}} Type 1: \\Electric vehicle\end{tabular}} & \multirow{2}{*}{$1$}                  & \multirow{2}{*}{$9$}                 & \multirow{2}{*}{$0$}                       & \multirow{2}{*}{$20$}                     & \multirow{2}{*}{$50$}           \\
		&                                       &                                      &                                            &                                           &                                 \\ \hline
		\multirow{2}{*}{\begin{tabular}[c]{@{}c@{}}Type 2:\\ EWH-morning\end{tabular}} & \multirow{2}{*}{$5$}                  & \multirow{2}{*}{$8$}                & \multirow{2}{*}{$0$}                      & \multirow{2}{*}{$12$}                     & \multirow{2}{*}{$9$}          \\
		&                                       &                                      &                                            &                                           &                                 \\ \hline
		\multirow{2}{*}{\begin{tabular}[c]{@{}c@{}}Type 3:\\ EWH-evening\end{tabular}}      & \multirow{2}{*}{$16$ }                  & \multirow{2}{*}{$19$}                 & \multirow{2}{*}{$0$}                       & \multirow{2}{*}{$12$}                     & \multirow{2}{*}{$9$}           \\
		&                                       &                                      &                                            &                                           &                                 \\ \hline
		\multirow{2}{*}{\begin{tabular}[c]{@{}c@{}}Type 4: \\Dryer\end{tabular}}      & \multirow{2}{*}{$9$}                  & \multirow{2}{*}{$21$}                 & \multirow{2}{*}{$0$}                       & \multirow{2}{*}{$2.95$}                     & \multirow{2}{*}{$2.95$}           \\
		&                                       &                                      &                                            &                                           &                                 \\ \hline
	\end{tabular} 
\end{table}

We assume that each user has one or multiple types of controllable loads. Details of the users' controllable loads are given in Table \ref{Table:Flexible_Load}. It is shown that the EV needs to be charged from 00:00 AM to 8:00 AM (time slots $1 \le n \le 9$) to receive the total energy of $50$ kWh during this period. Electrical water heater (EWH) is considered to be used either in the morning or evening  and  needs to receive $8.85$ kWh to warm $50$ gallons of water from 04:00 AM to 07:00 AM (time slots $5 \le n \le 8$) and/or from 02:00 PM to 05:00 PM (time slots $16 \le n \le 19$), respectively \cite{RWH}. 
Finally, the dryer can operate flexibly during 08:00 AM to 08:00 PM (time slots $9  \le n \le 21$) and consume $3.19$ kWh during this time \cite{Dryer}. We assume that user 1 has  controllable loads of Class 1, user 2 Classes 2 and  4, user 3 Classes 3 and 4, and finally user 4 Classes 2 and 3.

For the shared ESS, we consider sodium-sulfur batteries with maximum and minimum capacities of $\overline{S}=18$ kWh and $\underline{S}=0.1\overline{S}$, respectively, charging and discharging efficiencies of $\dot{\alpha}=\ddot{\alpha}=0.87$, and maximum charging and discharging rates of $\overline{C}=\overline{D}=0.15\overline{S}$ \cite{SS_Battery}. We also set $S_1=\underline{S}$. Last, we consider $\beta_m=0.25$, $\forall m$,   set the price of purchasing energy from the grid as $0.2$ $\$$/kW \cite{EMA}, and model the cost function as  $f_{mn}=0.2G_{mn}$. 

In the following, we provide  numerical examples to first show the energy cost saving resulting from the shared ESS, compared to the case of distributed ESSs. Next, we evaluate the performance of the three online algorithms. Finally, we highlight the impact of renewable energy diversity on the effectiveness of shared ESS in energy cost saving. 
\subsection{Shared versus Distributed ESSs}
In this subsection, we aim to show the effectiveness of the shared ESS over distributed ESSs (with the optimization problem given in (P2)) in energy cost saving. 
\begin{figure}[t!]
	\centering
	\includegraphics[width=8.5cm]{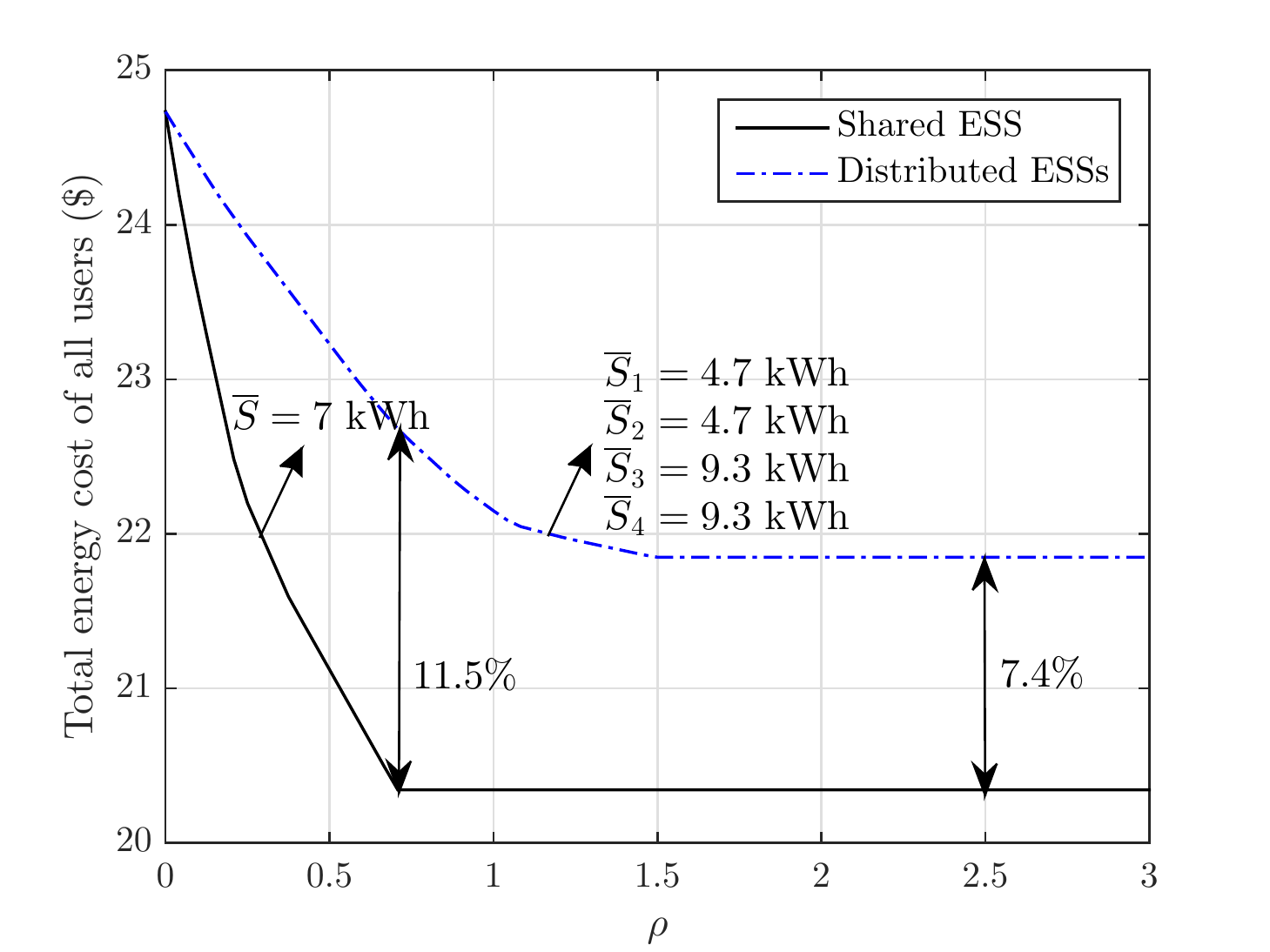} \vspace{-2mm}
	\caption{Total energy cost  from shared and distributed ESSs schemes.}\label{fig:SharedvsDist}	 
\end{figure}
\begin{figure}[t!]
	\centering
	\includegraphics[width=8.5cm]{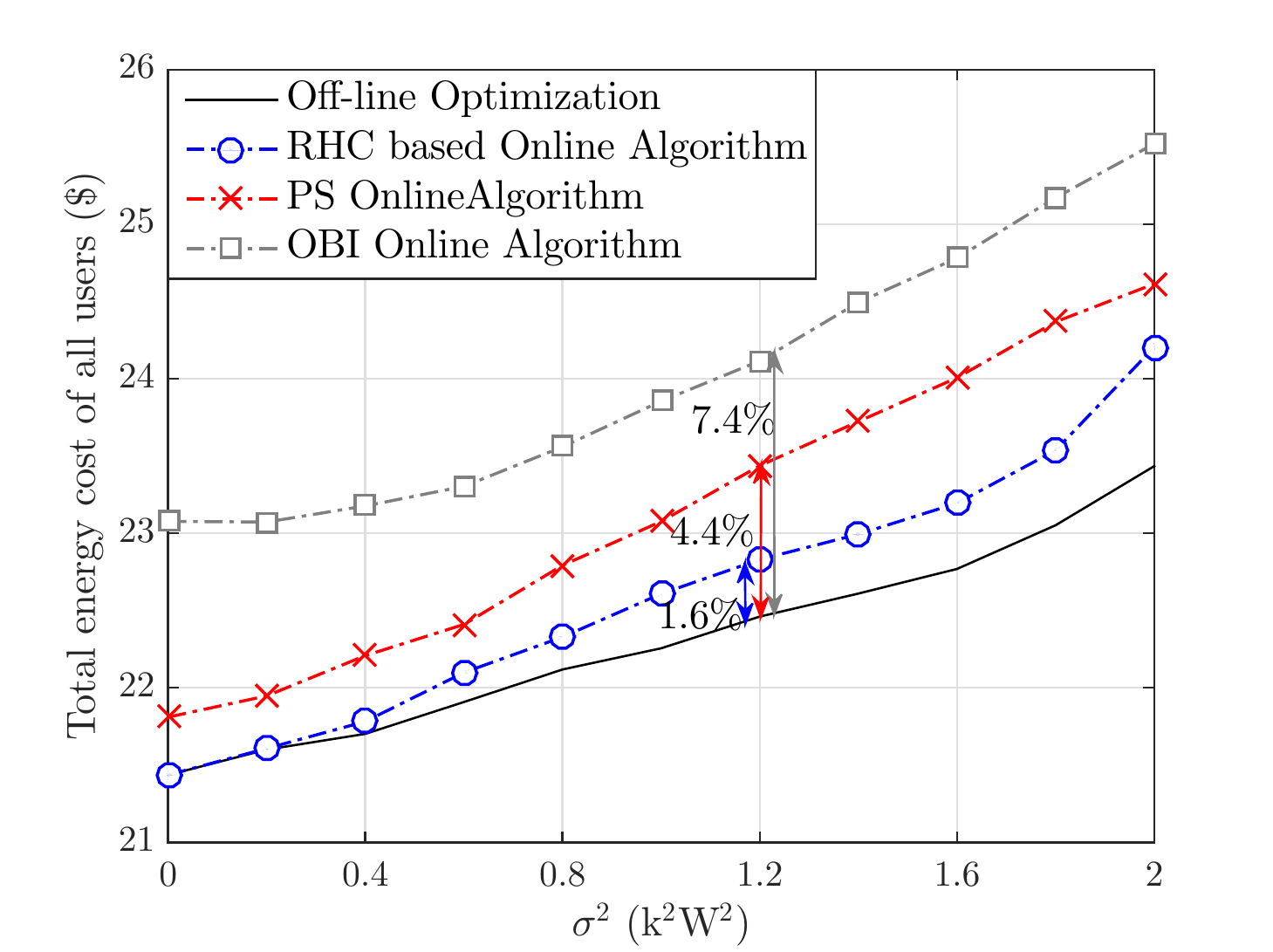} \vspace{-2mm}
	\caption{Total energy cost versus the variance of prediction errors.}\label{fig:Online} 
\end{figure}
The total energy cost of all users resulting from shared and distributed ESSs schemes over $\rho\overline{S}$ ($\rho\overline{S}_m$ for distributed ESSs) are shown in Fig. \ref{fig:SharedvsDist}. It is observed that the total energy cost of users decreases over ESS capacity, which is due to less waste in surplus energy. 
Furthermore, it is observed that users with a shared ESS can achieve a total energy cost target with a smaller ESS capacity as compared to the case of distributed ESSs. 
For instance, to achieve the total energy cost of $\$22$, the capacity of the shared ESS can be set as  $\overline{S}=7$ kWh. However, for the case of distributed ESSs, we need to set $\overline{S}_1=4.7$ kWh, $\overline{S}_2=4.7$ kWh, $\overline{S}_3=9.3$ kWh, and $\overline{S}_4=9.3$ kWh, where the overall capacity in this case is $28$ kWh. This shows that the shared ESS significantly reduces the overall ESS capacity requirement  by enabling energy sharing among users.  
In addition, the shared ESS can avoid renewable energy curtailments more effectively over the case of distributed ESSs, due to its higher capacity compared to each individual distributed ESS.  
\subsection{Performance Evaluation of Online Algorithms}\label{SubSec:Online}
In this subsection, we aim to evaluate the performance of the online algorithms in Section \ref{Sec:Online}, under unknown prediction errors.  
We assume that prediction errors $\delta_{mn}$'s follow independent and identical Gaussian distributions with zero mean and variance $\sigma_{mn}^2$. We then set $\sigma_{mn}^2=\sigma^2$, $\forall m \in {\cal M}$, $\forall n \in {\cal N}$. 

Fig. \ref{fig:Online} shows the average total energy  cost versus the prediction error variance $\sigma^2$. First, the off-line optimization is observed to outperform the three online algorithms, since it is under the ideal assumption that renewable energy generation/load are completely
known. It is also observed that RHC based online algorithm achieves its cost very close to the minimum cost by off-line optimization, and also outperforms over PS and OBF online algorithms. This is expected, since RHC deploys Algorithm 1 
to solve the energy management problem at each time slot $n$, by iteratively exchanging information between users and the central controller and exploiting the future predictable values of net energy profiles.
Therefore, the capacity of the shared ESS and the flexibility of controllable loads are fully utilized and the resulting energy cost is lower compared to the other two alternative online algorithms that make decisions only based on the current state of the system and limited information received from users. However, the proposed PS and OBF online algorithms still perform close to the optimal off-line solution with performance losses of  $4.4$$\%$ and $7.4$$\%$, respectively, in the noisy environment with
$\sigma^2=1.2$ k$^2$W$^2$. 
\subsection{Impact of Renewable Energy Diversity}

\begin{figure}[t!]
	\centering
	\includegraphics[width=8.5cm]{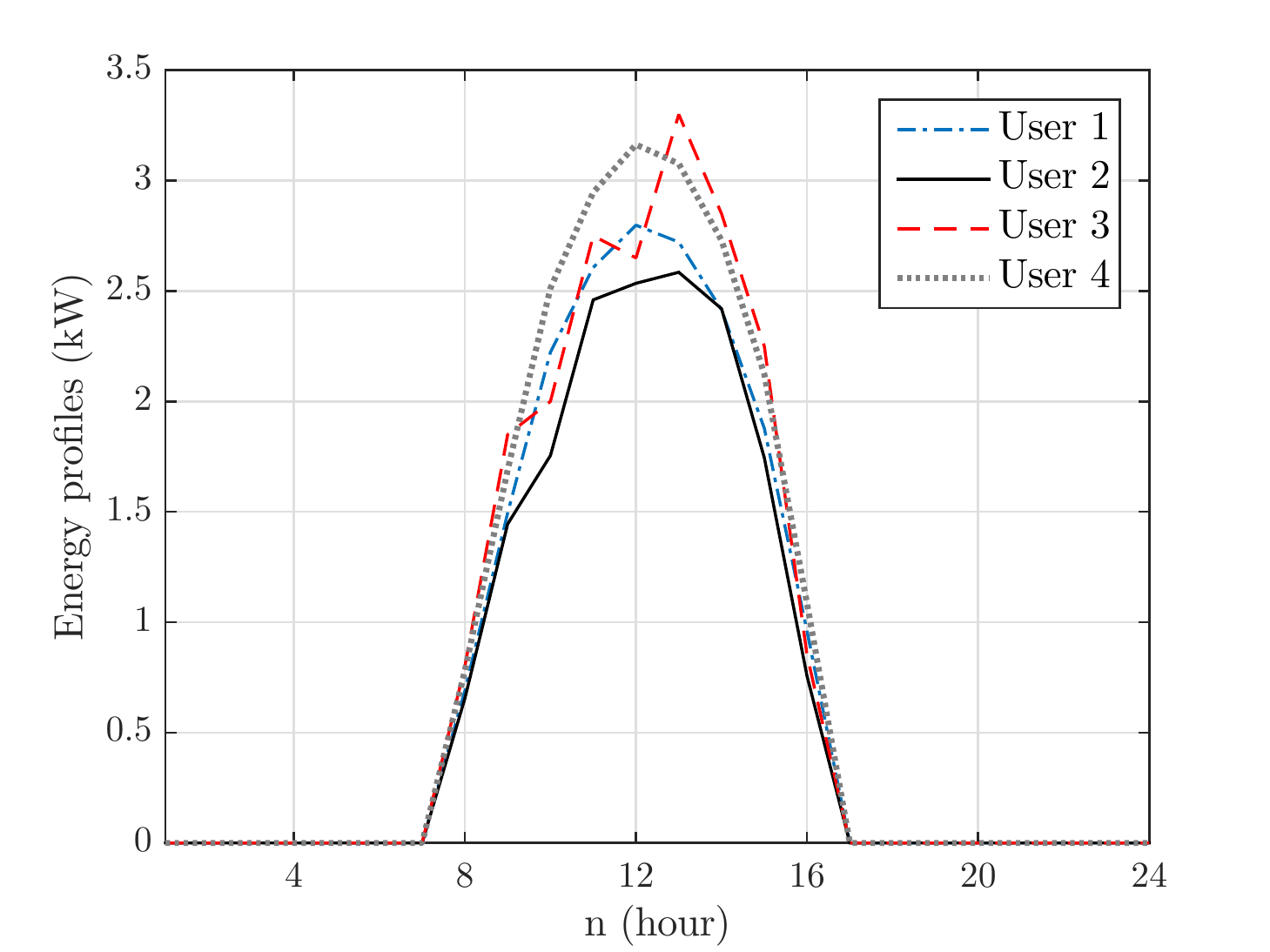} 
	\caption{Solar energy profiles of users in the low diversity case.}\label{fig:OnlySolar} 
\end{figure}

\begin{figure}[t!]
	\centering
	\includegraphics[width=12cm]{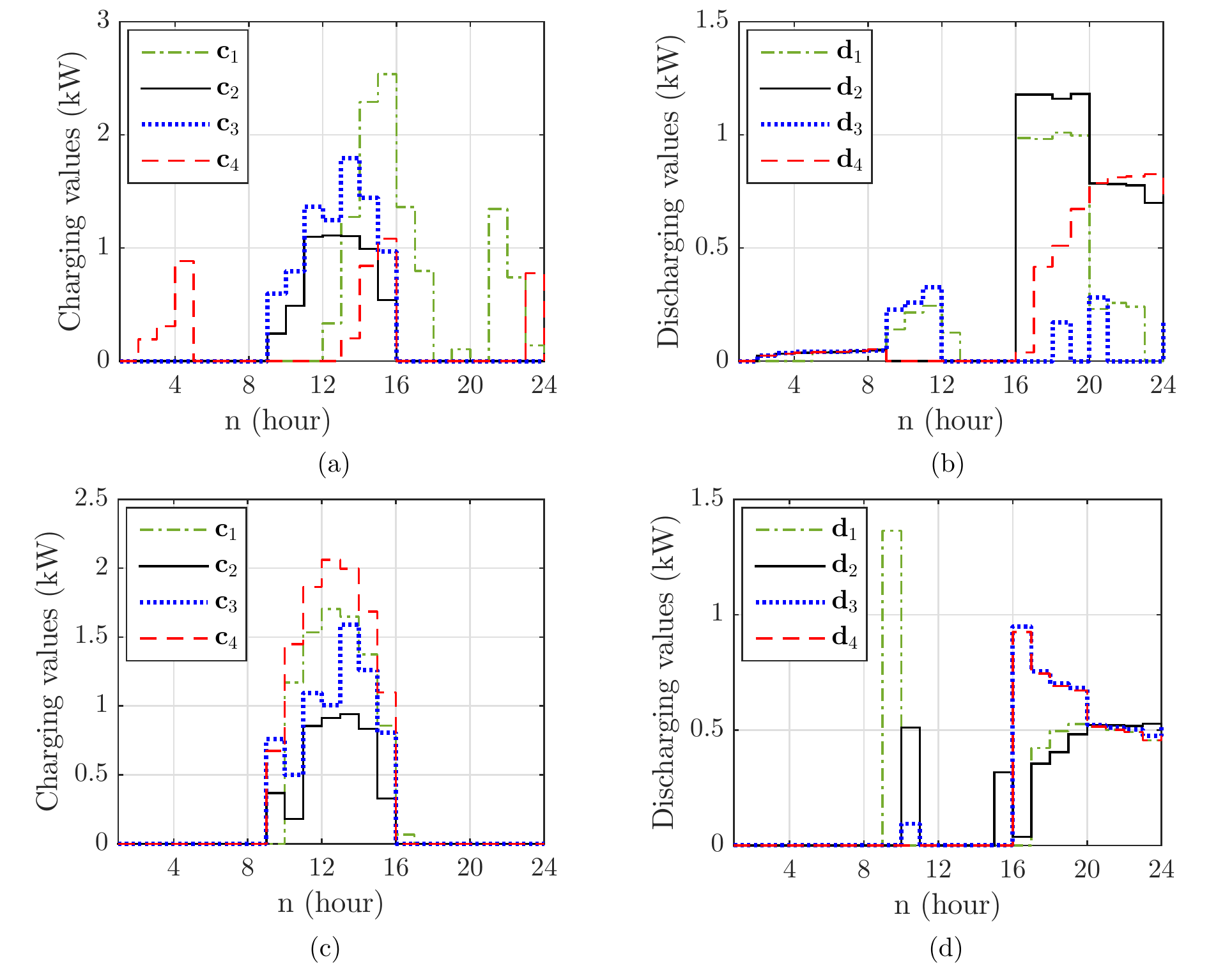} \vspace{-3mm}  
	\caption{Charging and discharging values given $\overline{S}=18$ kW: (a) and (b) for high diversity; (c) and (d) for low diversity.}\label{fig:Charging&Discharging} 
\end{figure}

By keeping the fix load profiles of users unchanged, we consider that users 1 and 4, similar to users 2 and 3, only have solar energy sources. Renewable energy profiles of all four users in this case are shown in Fig. \ref{fig:OnlySolar}. 
The goal is to compare the effectiveness of the shared ESS in energy cost saving in two different setups of only solar (low diversity) and diverse renewable energy sources of solar, wind, or both (high diversity).
\begin{figure}[t!]
	\centering
	\includegraphics[width=11.25cm]{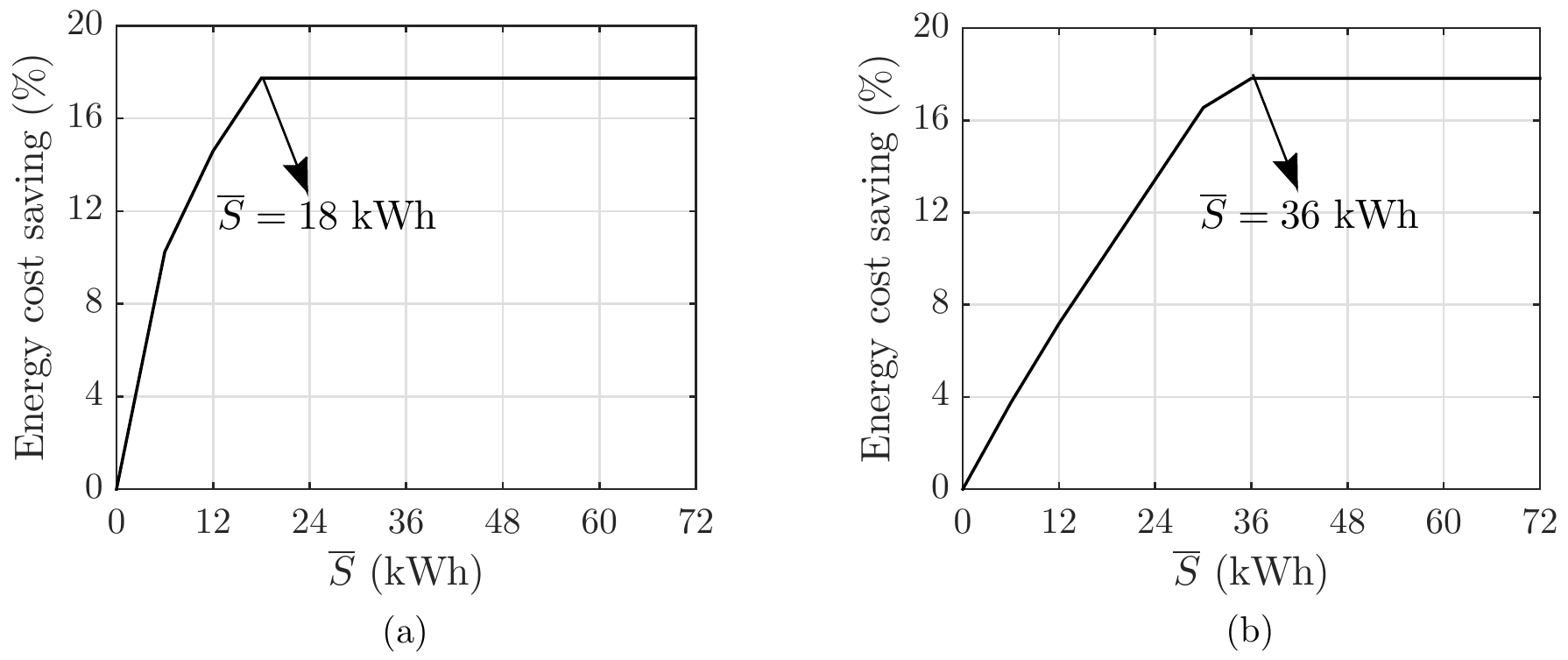} \vspace{-2mm}
	\caption{Energy cost saving: (a) High diversity (wind and solar energy generators), (b) Low diversity (only solar energy generators).}\label{fig:Diversity}
\end{figure} 
The total energy cost saving of all users for the two cases of high and low renewable energy diversities are shown in Fig. \ref{fig:Diversity}-a and \ref{fig:Diversity}-b, respectively. It is observed that energy cost saving in the high diversity case in Fig. \ref{fig:Diversity}-a remains unchanged over the  shared ESS capacity increase for $\overline{S}\ge 18$ kWh, while in \ref{fig:Diversity}-b, it happens for $\overline{S}\ge 36$ kWh, which shows that the highest achievable energy cost saving is attainable in significantly lower ESS capacity when the diversity is high.
This is because when diversity is high, it is more likely that the energy surplus/deficit in users' renewable energy profiles do not happen at the same time. 
In this case, the surplus energy of some users can compensate the energy deficit in others and charging/discharging to/from the shared ESS do not happen concurrently, as validated in Figs. \ref{fig:Charging&Discharging}-a and \ref{fig:Charging&Discharging}-b.  
This in contrast to the low diversity case in which charging and discharging values happen almost at the same time slots, as shown in Figs. \ref{fig:Charging&Discharging}-c and \ref{fig:Charging&Discharging}-d.
%
\section{Conclusion}\label{Sec_Concl}
In this paper, we address the energy management problem of a system of multiple renewable energy integrated users sharing a common ESS. First, we propose an algorithm for the optimal off-line energy management that can be implemented in a distributed manner and by exchanging limited amount of information among users and the central controller. Next, we propose three online algorithms that differ in complexity, information sharing, and performance. We discuss each algorithm in detail and evaluate their performance via simulations, using a practical system setup. We also make comparison with the case of distributed ESSs, where each user owns its relatively smaller-scale ESS, which is not shared  with others. The simulation results show that the shared ESS can potentially decrease the total energy cost of all users compared to the case of distributed ESSs by enabling energy sharing among them.

\end{document}